\documentclass[twocolumn,superscriptaddress,floatfix]{revtex4-1}

\usepackage{amsfonts,amsmath,graphicx,amssymb,bm}
\usepackage[pdftex,plainpages=false,colorlinks=true,linkcolor=blue, citecolor=blue, urlcolor=blue]{hyperref}

\pdfoutput=1

\usepackage[x11names,svgnames]{xcolor}

\newcommand\er{\mathrm{e}}
\newcommand\dr{\mathrm{d}}
\newcommand\ev{\mathbf{e}}

\newcommand\kv{\mathbf{k}}
\newcommand\rv{\mathbf{r}}

\newcommand\kvt{\mathbf{\tilde k}}
\newcommand\Gv{\mathbf{G}}

\newcommand\tv{\mathbf{t}}

\newcommand\thetav{\bm{\theta}}
\newcommand\Gammav{\bm{\Gamma}}
\newcommand\Sigmav{\bm{\Sigma}}
\newcommand\epsilonv{\bm{\epsilon}}
\newcommand\om{\omega}

\newcommand\Tr{\,\mathrm{Tr}\,}

\newcommand\Hc{\mathrm{H.c.}}

\renewcommand\Re{\,\mathrm{Re}\,}
\renewcommand\Im{\,\mathrm{Im}\,}
\newcommand\Ccal{\mathcal{C}}

\begin{document}
\title{$d$-wave superconductivity in coupled ladders}

\author{J. P. L. Faye}
\affiliation{D\'epartment de physique and RQMP, Universit\'e de Sherbrooke, Sherbrooke, Qu\'ebec, Canada J1K 2R1}

\author{S. R. Hassan}
\affiliation{The Institute of Mathematical Sciences, C.I.T. Campus, Chennai 600 113, India}

\author{P.V. Sriluckshmy}
\affiliation{The Institute of Mathematical Sciences, C.I.T. Campus, Chennai 600 113, India}

\author{G. Baskaran}
\affiliation{The Institute of Mathematical Sciences, C.I.T. Campus, Chennai 600 113, India}
\affiliation{Perimeter Institute of Theoretical Physics, Waterloo, Ontario, Canada.}

\author{D. S\'en\'echal}
\affiliation{D\'epartment de physique and RQMP, Universit\'e de Sherbrooke, Sherbrooke, Qu\'ebec, Canada J1K 2R1}

\date{\today}

\begin{abstract}
We study the one-band Hubbard model on the trellis lattice, a two-dimensional frustrated lattice of coupled two-leg ladders, with hopping amplitude $t$ within ladders and $t'$ between ladders.
For large $U/t$ this is a model for the cuprate Sr$_{14-x}$Ca$_x$Cu$_{24}$O$_{41}$.
We investigate the phase diagram as a function of doping for $U=10t$ using two quantum cluster methods:
The variational cluster approximation (VCA), with clusters of sizes 8 and 12, and Cellular dynamical mean field theory (CDMFT), both at zero temperature.
Both methods predict a superconducting dome, ending at roughly 20\% doping in VCA and 15\% in CDMFT.
In VCA, the superconducting order parameter is complex in a range of doping centered around 10\%, corresponding to bulk chiral, $T$-violating superconductivity. However, the CDMFT solution is not chiral.
We find evidence for a migration of the Cooper pairs from the inter-ladder region towards the plaquettes as doping is increased.
\end{abstract}
\maketitle

\section{Introduction}
Inspired by the discovery of high-$T_c$ super\-conductivity, Dagotto {\it et al.} predicted a superconducting phase in a theoretical model consisting of weakly-coupled, quasi-one-dimensional ladders.
This model exhibits a spin-gap and $d$-wave hole-pair formation away from half filling~\cite{Dagotto:1988,Dagotto:1992,Barnes:1993}.
This prediction was realized in the copper oxide-based ladder material Sr$_{14-x}$Ca$_x\rm Cu_{24}O_{41}$ with hole doping. 
At $x=13.6$, the critical temperatures under pressures of 3 GPa and 4.5 GPa are $T_c = 12$K and $T_c = 9$K, respectively~\cite{Uehara:1996}.
At $x=11.5$, a superconducting dome is seen as a function of pressure~\cite{Nagata:1997}.
Recently, the critical temperature of the $x=11$ compound $\rm Sr_{3}Ca_{11}Cu_{24}O_{41}$ has been doubled, from $12$K to $24$K,  by applying a weak uniaxial pressure of 0.06 GPa~\cite{Mohan-Radheep:2013kq}.

Many theoretical studies have been reported on the single-ladder Hubbard model, with and without doping, using a variety of methods: Exact diagonalizations~\cite{ Dagotto:1992,Troyer:1996fk}, density-matrix renormalization group~\cite{Noack:1994}, resonating-valence-bond (RVB) mean-field theory~\cite{Gopalan:1994}, bosonization~\cite{Balents:1996} and quantum Monte Carlo (QMC)~\cite{Kuroki:1996, Dahm:1997}.
A consistent result from those studies is the emergence of $d$-wave superconducting correlations in the double leg ladder upon doping. 
Coupled ladders described by the trellis lattice have been investigated using the Fluctuation Exchange (FLEX) method, confirming the possibility of $d$-wave superconductivity at half-filling~\cite{Kontani:1998}.

In this paper, we report on a theoretical study of superconductivity in the one-band Hubbard model on the trellis lattice away from half-filling, at zero temperature. We use the Variational Cluster Approximation (VCA)~\cite{Dahnken:2004} and Cellular Dynamical Mean-Field Theory (CDMFT)~\cite{Lichtenstein:2000vn,Kotliar:2001}. For the range of on-site repulsion studied, superconductivity does not occur at half-filling, but a superconducting dome appears upon doping. Moreover, the superconducting order parameter computed from VCA becomes complex in a range of doping centered around 10\%, thus breaking time-reversal symmetry.
The energy gain from the chiral nature of superconductivity is small, at best $\frac1{15}$ of the condensation energy, and the chiral solution is not found with CDMFT.

This paper is organized as follows. In Section~\ref{sec:bcs} the model is presented, as well as the structure of singlet superconductivity in the BCS approximation. In Section~\ref{sec:vca} the VCA technique is summarized and the results of its application are presented; this is the main part of the paper. In Section~\ref{sec:cdmft} CDMFT is applied in order to confirm by an independent method the occurrence of superconductivity. We add a short discussion and conclude in Section~\ref{sec:discussion}.

\begin{figure}[tbh]
\includegraphics[scale=0.8]{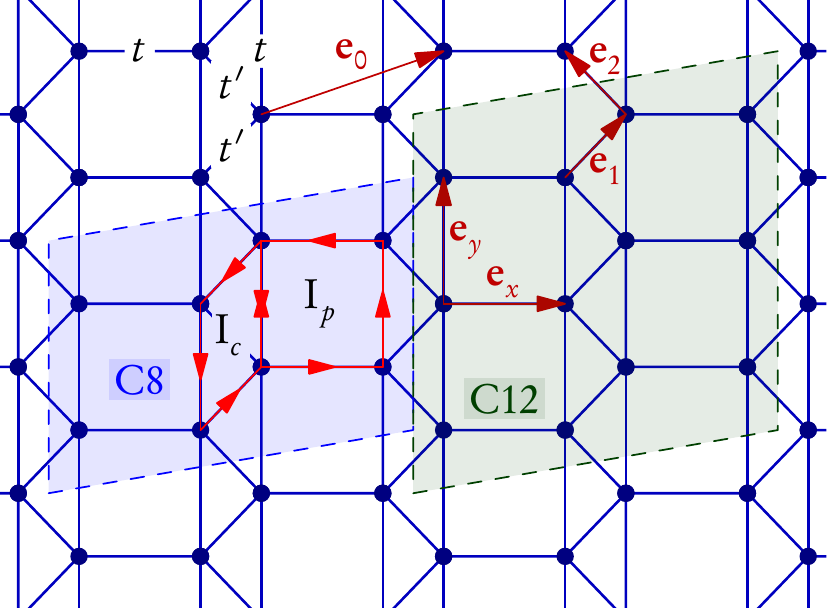}
\caption{(Color online) Model for the ladder cuprate Sr$_{14-x}$Ca$_x\rm Cu_{24}O_{41}$ on the trellis lattice. 
The shaded areas are the clusters used in VCA, labeled C8 and C12.}      
\label{fig:cluster}
\end{figure}

\section{Model and mean-field representation}\label{sec:bcs}
The Hamiltonian of the one-band, repulsive Hubbard model on the trellis lattice is
\begin{equation}\label{eq:H}
 H = \sum_{i,j,\sigma} t_{ij}c^{\dag}_{i \sigma} c_{j \sigma} +U \sum_in_{i\uparrow}n_{i\downarrow}
 -\mu\sum_i (n_{i\uparrow}+n_{i\downarrow})
\end{equation}
where $c_{i\sigma}$ annihilates an electron of spin $\sigma$ at site $i$, $t_{ij}=t_{ji}$ is the hopping amplitude between sites $i$ and $j$, $n_{i\sigma} = c^{\dag}_{i\sigma}c_{i\sigma}$ is the number operator at site $i$ and $U$ the on-site Coulomb repulsion. The density of electrons is controlled by the chemical potential $\mu$.
The only nonzero hopping terms are indicated by inter-site links on Fig.~\ref{fig:cluster}, with $t_{ij}=t$ on the ladder plaquettes, and $t_{ij}=t'$ between ladders.
Model~\eqref{eq:H} has two bands: the unit cell contains two orbitals, separated by $\ev_x$ on the figure.
The vectors $\ev_0$ and $\ev_y$ on Fig.~\ref{fig:cluster} define a basis for the lattice.
For convenience, we will define the Brillouin zone exactly like on the graphene lattice.

This model offers an approximate description of $\rm Sr_{3}Ca_{11}Cu_{24}O_{41}$, wherein
each site represents a copper atom. In the actual material, oxygen atoms are located midway between copper atoms on the links of each square plaquette.
The relation between hole doping $\delta$ (the electron density is $n=1-\delta$) and Ca doping $x$ in the material is not simple, as $\delta$ is also affected by pressure.
Throughout this paper we will set $t=1$ and $t'=0.15$; this ratio $t'/t$ is taken from band structure calculations~\cite{Arai:1997}.
The value of $U/t$ will be set to 10 in most VCA and CDMFT computations.

\subsection{Superconductivity}

What form can superconductivity take in such a model?
To answer this question, let us first provide a description of the superconducting order-parameter at the mean-field level.
It is then convenient to adopt a Nambu description, with the multiplet of destruction/creation operators
\begin{equation}\label{eq:nambu}
\Ccal_\kv = \big(c_{1\uparrow}(\kv), c_{2\uparrow}(\kv), c^\dagger_{1\downarrow}(-\kv), c^\dagger_{2\downarrow}(-\kv) \big)
\end{equation}
where the first index of each operator is a sublattice index, distinguishing the left and right sites of each rung.
This array of operators is used in a matrix description of the non-interacting, mean-field Hamiltonian
\begin{equation}\label{eq:H_BCS}
H_{\rm BCS} = \sum_\kv \Ccal_\kv^\dagger H_\kv \Ccal_\kv
\end{equation}
with the $4\times4$ Hermitian matrix
\begin{widetext}
\begin{equation}
H_\kv = 
\begin{pmatrix}
-2\cos(\kv\cdot\ev_y) - \mu & \gamma_\kv & \theta_\kv & \eta_\kv \\
\gamma^*_\kv & -2\cos(\kv\cdot\ev_y) - \mu & \eta_{-\kv} & \theta_\kv \\
\theta^*_\kv & \eta^*_{-\kv} & 2\cos(\kv\cdot\ev_y)+\mu & -\gamma_\kv \\
\eta^*_\kv & \theta^*_\kv & -\gamma^*_\kv & 2\cos(\kv\cdot\ev_y) + \mu
\end{pmatrix}
\end{equation}
\end{widetext}
with $\gamma_\kv = -\er^{-i\kv\cdot\ev_x} - t'\er^{-i\kv\cdot\ev_1}- t'\er^{-i\kv\cdot\ev_2}$.
This is the most general form for singlet superconductivity.
If we assume only nearest-neighbor pairing with amplitudes $D_x$, $-D_y$ and $D_{1,2}$ in the directions $\ev_x$, $\ev_y$ and $\ev_{1,2}$ respectively, the anomalous terms of that matrix are
\begin{eqnarray}
\theta_\kv &=& -D_y\cos(\kv\cdot\ev_y) \notag \\
\eta_\kv &=& D_x \er^{i\kv\cdot\ev_x} +D_1\er^{i\kv\cdot\ev_1}+D_2\er^{i\kv\cdot\ev_2}
\end{eqnarray}
The choice of sign for $D_y$ reflects our anticipation of $d$-wave superconductivity on the plaquettes.

If the superconductor is time-reversal ($T$) invariant, the components of $H_\kv$ satisfy the relation $H_{-\kv} = H^*_\kv$.
This implies that the amplitudes $D_{x,y,1,2}$ defined above are all real.
On the other hand, if any one of them is complex, the superconductor breaks time-reversal invariance.

Let us stress that we are not performing a true mean-field computation here: there is no factorization of the interaction, no self-consistent procedure, etc. Indeed, the Hubbard model, with its local repulsion, is not amenable to a self-consistent (BCS) mean-field computation of $d$-wave superconductivity. We are simply illustrating the form that superconductivity can take in this model, in order to compare with the complete variational or self-consistent computations reported on in Section~\ref{sec:vca}. 

\begin{figure}
\includegraphics[width=\hsize]{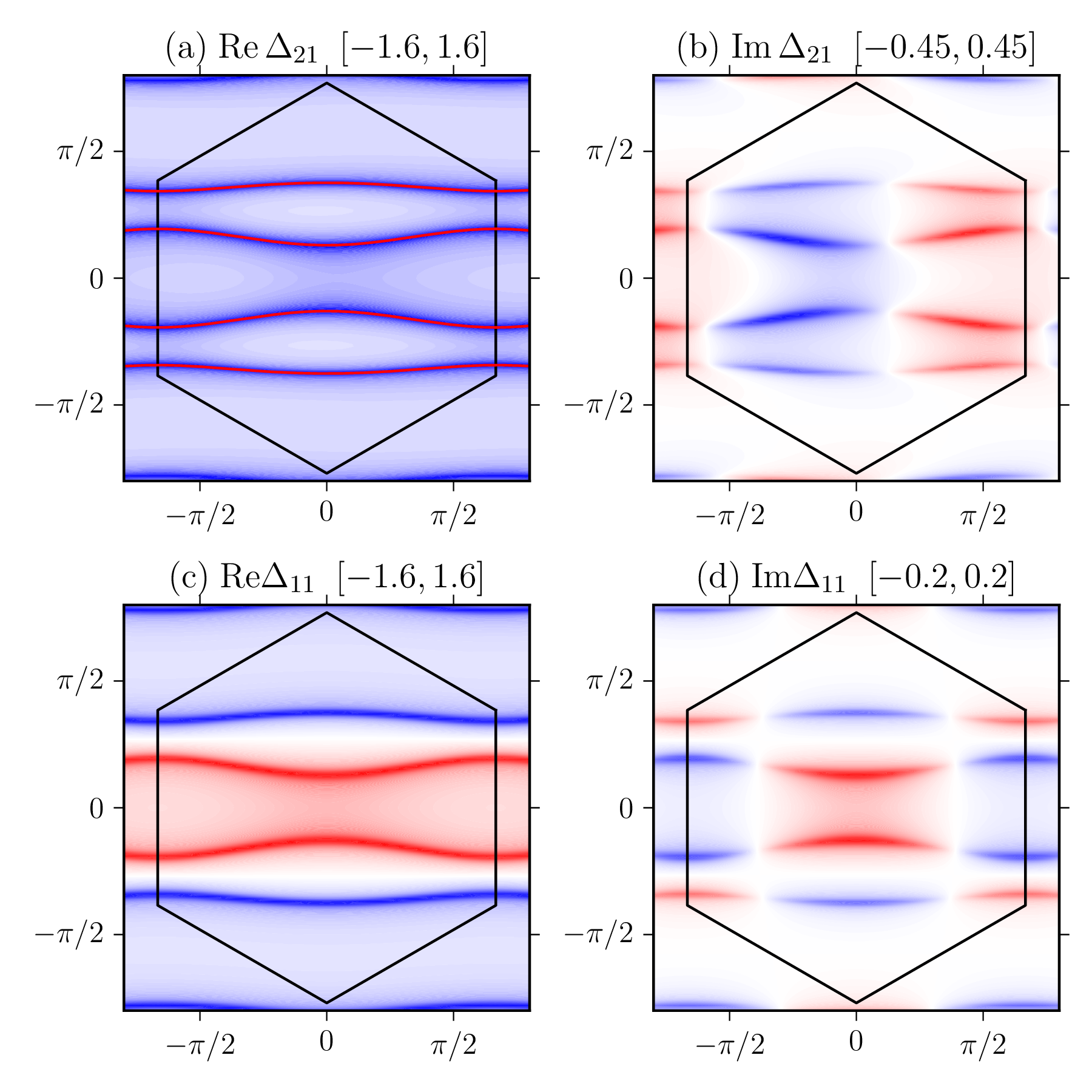}
\caption{(Color online) Superconducting order parameter $\Delta_{ab}(\kv)$ for the BCS Hamiltonian \eqref{eq:H_BCS}, as a function of wave-vector, with parameters $D_x=0.09$, $D_y=0.03$ and $D_1=D_2=-0.01+0.008i$. Doping is set at 10\%. The Brillouin zone is indicated. Panels (a) and (b) show the real and imaginary parts of the rung component (different sublattices), whereas panels (c) and (d) show the leg component (same sublattice). Red means negative, blue positive and the range is indicated on top of each panel.
The normal state Fermi surface is shown in red on panel (a).} 
\label{fig:gorkov_bcs}
\end{figure}

\subsection{Order parameter}

The most general way to represent superconducting order is to plot the momentum-dependent order parameter $\Delta_{ab}(\kv)$,
defined as the integral over frequency of the Gorkov function (the anomalous part of the Green function):
\begin{equation}
\Delta_{ab}(\kv) = \int_{-\infty}^\infty \frac{\dr\omega}{2\pi} F_{ab}(\kv,i\omega)
\end{equation}
Here $(a,b)$ are sublattice indices taking two possible values, associated with the left and right legs of the ladder.
The Gorkov function $F_{ab}$ is the top-right block of the Nambu Green function $G_{\mu\nu}(\kv,\om)$ defined as follows at zero temperature:
\begin{multline}
G_{\mu\nu}(\kv,\om) = \langle\Omega|\Ccal_{\mu}(\kv)\frac1{\om-H+E_0}\Ccal^\dagger_{\nu}(\kv)|\Omega\rangle \\
 + \langle\Omega|\Ccal^\dagger_{\nu}(\kv)\frac1{\om+H-E_0}\Ccal_{\mu}(\kv)|\Omega\rangle~~,
\end{multline}
where $\om$ is a complex-valued frequency, $|\Omega\rangle$ is the many-body ground state and $E_0$ the ground state energy.
For a two-band model, $F_{ab} = G_{a,b+2}$.
In the special case of the non-interacting BCS Hamiltonian \eqref{eq:H_BCS}, the Nambu Green function is
\begin{equation}
G_{\mu\nu}(\kv,\om) = \left(\frac1{\om - H_\kv}\right)_{\mu\nu}
\end{equation} 

In order to connect with the more familiar one-band BCS theory, let us point out that in that case the matrix $H_\kv$ has the simpler form
\begin{equation}
H_\kv = \begin{pmatrix}
\epsilon_\kv - \mu & \theta_\kv \\
\theta^*_\kv & -\epsilon_\kv + \mu
\end{pmatrix}
\end{equation}
where $\epsilon_\kv$ is the dispersion relation and $\theta_\kv$ the gap function.
The order parameter is then simply
\begin{equation}
\Delta_\kv = -\frac{\theta_\kv}{2\sqrt{\epsilon_\kv^2+|\theta_\kv|^2}}
\end{equation}

Figure \ref{fig:gorkov_bcs} illustrates the superconducting order parameter $\Delta_{ab}(\kv)$ in a particular case of Model \eqref{eq:H_BCS}.
This will later be compared to a solution, obtained through VCA, that contains correlation effects.
We have chosen superconducting amplitudes $(D_x,D_y,D_1,D_2)$ that break time reversal slightly and match local order parameters of an actual VCA solution described later on.
The fact that $\Im\Delta_{12}(\kv) \ne -\Im\Delta_{12}(-\kv)$ and $\Im\Delta_{11}(\kv) \ne 0$ is a visual signature of time-reversal breaking.

\section{The Variational Cluster Approximation}\label{sec:vca}
We use the Variational Cluster Approximation (VCA)~\cite{Dahnken:2004} to investigate the zero-temperature phase diagram of Model~\eqref{eq:H}, more specifically the existence of $d$-wave superconductivity upon hole doping for several values of $U$.
VCA -- also called VCPT in its early days -- has been used to study the emergence of $d$-wave superconductivity in a simple description of the high-$T_c$ cuprates based on the Hubbard model~\cite{Senechal:2005,Aichhorn:2006rt}.
It is based on Potthoff's self-energy functional approach~\cite{Potthoff:2003b}.
For a review, see Ref.~\cite{Potthoff:2012}.

\subsection{Description of the method}
In VCA, we must distinguish between the original Hamiltonian $H$, defined on the infinite lattice, and a reference Hamiltonian $H'$, defined on a small cluster of atoms. $H'$ is a restriction of $H$ to the cluster, except that a finite number of Weiss fields may be added to it, in order to probe certain broken symmetries. Any one-body term can also be added to $H'$.
The electron self-energy $\Sigmav(\omega)$ associated with $H'$ is used as a variational self-energy, in order to construct the Potthoff self-energy functional:
\begin{multline}\label{eq:omega3}
 \Omega[\Sigmav(h)]=\Omega'[\Sigmav(h)]\\ +\Tr\ln[-(\Gv^{-1}_0 -\Sigmav(h))^{-1}]
 -\Tr\ln(-\Gv'(h))
\end{multline}
where $\Gv'$ is the physical Green function of the cluster, $\Gv_0$ is the non interacting Green function of the original model and $h$ denotes collectively the coefficients of all the adjustable one-body terms added to $H'$ acting as variational parameters.
The symbol $\Tr$ stands for a functional trace, i.e., a sum over all degrees of freedom (e.g. momenta or sites) and frequencies.
$\Omega'$ is the ground state energy (chemical potential included) of the cluster which, along with the associated Green function $\Gv'$, is computed numerically, in our case via the exact diagonalization method at zero temperature.

Eq.~\eqref{eq:omega3} provides us with an exact, non-perturbative value of the Potthoff functional $\Omega[\Sigma(h)]$, albeit 
on a restricted space of self-energies $\Sigma(h)$ which are the physical self-energies of the reference Hamiltonian $H'$.
Expression \eqref{eq:omega3} is computed numerically in order to look for stationary points of that functional, for instance via a Newton or quasi-Newton method.
The resulting value of $h$ defines the best possible self-energy $\Sigmav(\omega)$ for that parameter set; the latter is then combined with $\Gv_0$ to form an approximate Green function $\Gv$ for the original Hamiltonian $H$, from which any one-body quantity, for instance the order-parameters associated with broken symmetries, can be computed.

When confronted with competing solutions, i.e., different stationary points of $\Omega[\Sigma(h)]$ or points obtained via different sets of Weiss fields, the one with the lowest value of the Potthoff functional is selected, as $\Omega$ is an approximation of the exact free energy of the infinite system. 
VCA retains the correlated character of the model, since the local interaction is not factorized. 
The approximation may be controlled in principle by varying the size of the cluster and the number of variational parameters used.

In this work we use the clusters labeled C8 and C12 illustrated on Fig.~\ref{fig:cluster}.
It is important to test more than one cluster, as there will be some variance in numerical results among different clusters and robust characteristics need to be identified.
Larger clusters will generally lead to smaller values of the order parameter, because of in-cluster order parameter fluctuations.

\begin{figure} 
\includegraphics[width=\hsize]{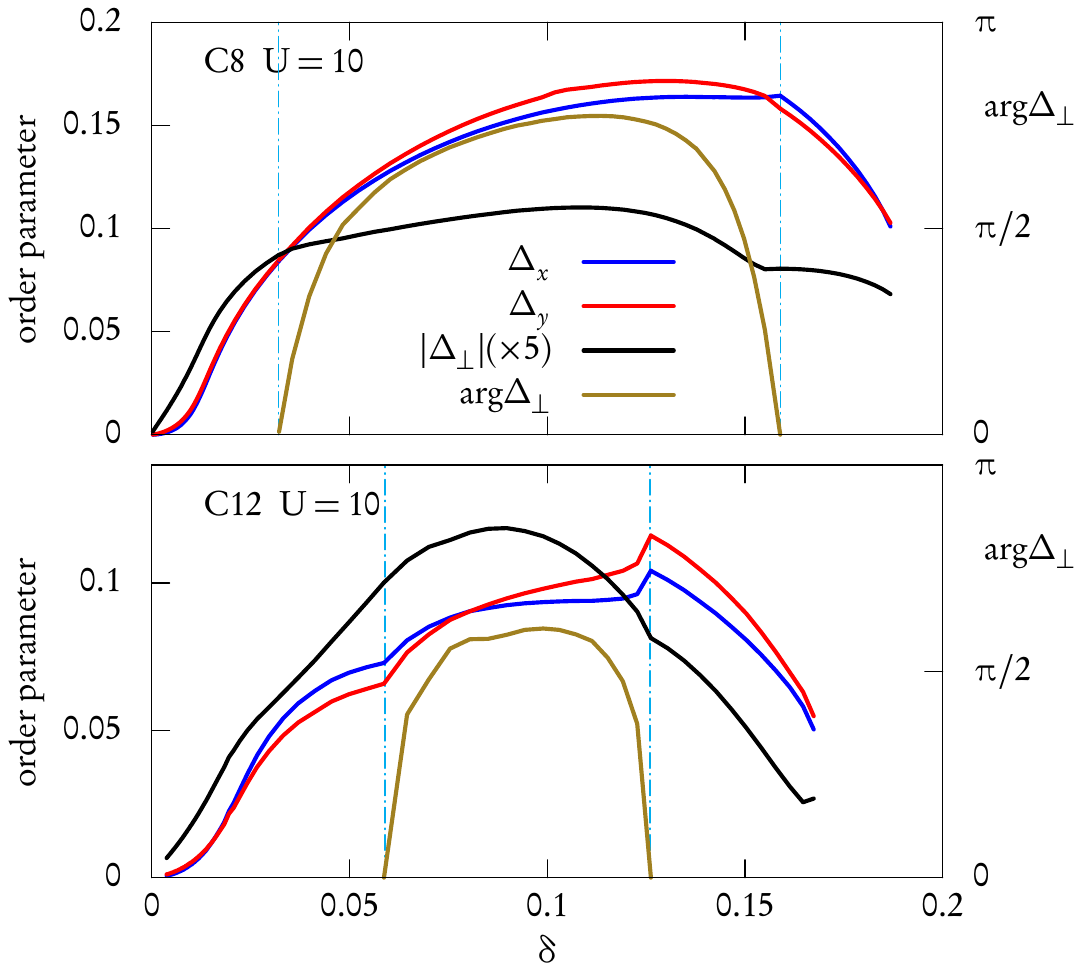}
\caption{(Color online) Pairing order parameters $(\Delta^x,\Delta^y,\Delta^\bot)$ as a function of doping $\delta$ for the two clusters used in VCA, at $U=10$. The complex phase of $\Delta^\bot$ can be read on the right axis.
The $T$-breaking solution ($\arg\Delta_\bot\ne0$) exists in a finite range of doping.} 
\label{fig:U10_C8-C12}
\end{figure}

\subsection{Superconductivity}
In VCA the possible presence of $d$-wave superconductivity is probed by adding to the cluster Hamiltonian $H'$ pairing operators. These may be expressed in terms of the singlet pairing operators
$\hat\Delta_{ij} = c_{i\uparrow} c_{j\downarrow} - c_{i\downarrow} c_{j\uparrow} = \hat\Delta_{ji}$
as follows:
\begin{align}\label{eq:OP}
&\hat\Delta_x = \sum_{\rv_i-\rv_j = \ev_x} \hat\Delta_{ij} \qquad\qquad 
\hat\Delta_y = -\sum_{\rv_i-\rv_j = \ev_y} \hat\Delta_{ij} \notag\\
&\hat\Delta_1 = \sum_{\rv_i-\rv_j = \ev_1} \hat\Delta_{ij} \qquad\qquad
\hat\Delta_2 = \sum_{\rv_i-\rv_j = \ev_2} \hat\Delta_{ij}
\end{align}
where the unit vectors $\ev_{x,y,1,2}$ are defined on Fig.~\ref{fig:cluster}.
$\hat\Delta_x$ is the sum of all pairing operators along rungs, $\hat\Delta_y$ is the sum of all pairing operators along legs, and $\hat\Delta_{1,2}$ are the sum of pairing operators between the ladders.
In practice, the Nambu description \eqref{eq:nambu} is used: A particle-hole transformation is applied to spin-down operators, giving the above pairing operators the appearance of hopping terms.

The Weiss Hamiltonian added to the cluster Hamiltonian $H'$ takes the form
\begin{equation}
H'_{\rm sc} = D_x \hat\Delta_x + D_y \hat\Delta_y + D_1 \hat\Delta_1 + D_2 \hat\Delta_2 + \Hc 
\end{equation}
where the coefficients $(D_x,D_y,D_1,D_2)$, the so-called \textit{Weiss fields}, are variational parameters, adjusted so as to make the Potthoff functional \eqref{eq:omega3} stationary (in practice, minimum).
The minus sign in front of $\hat\Delta_y$ in \eqref{eq:OP} means that we anticipate $d$-wave symmetry on the plaquette, i.e., we anticipate $D_x$ and $D_y$ to have the same sign, which is indeed what we find numerically.

In principle, the Weiss fields may be complex-valued, as $H'_{\rm sc}$ remains Hermitian anyway. 
However, we choose $D_x$ to be real so as to fix the overall phase.
Then $D_y$, $D_1$ and $D_2$ can be complex. In practice, in order to limit the number of variational parameters, we assume that $D_y$ is real and that $\Re D_2=\pm\Re D_1$ and $\Im D_2=\pm\Im D_1$. We found that the lowest minima of the Potthoff functional have $D_1=D_2\equiv D_\bot$, and in the rest of this paper we will accordingly define $\hat\Delta_\bot = \hat\Delta_1 + \hat\Delta_2$, for a total of 4 variational parameters: $D_x$, $D_y$, $\Re D_\bot$ and $\Im D_\bot$.

Figure~\ref{fig:U10_C8-C12} shows the order parameters $\Delta_\alpha=\langle\hat\Delta_\alpha\rangle+\langle\hat\Delta^\dagger_\alpha\rangle$, for $\alpha=x,y,\bot$, computed from the VCA Green function, as a function of hole doping, for $U=10$ and the two clusters shown on Fig.~\ref{fig:cluster}. 
The inter-ladder order parameter $\Delta_\bot$ is complex over a range of doping: its modulus is plotted, along with its phase (right vertical axis).
The superconducting dome has a maximum between 10\% and 15\% (depending on the cluster) and  ends at about 20\% doping. 
It falls to zero exactly at half-filling.
The inter-ladder order parameter $\Delta_\bot$ is noticeably smaller than the ladder order parameters $\Delta_x$ and $\Delta_y$, but this is roughly in line with the ratio $t'/t=0.15$.
The rung and leg order parameters ($\Delta_x$ and $\Delta_y$) also have imaginary parts whenever $\Delta_\bot$ has one, but they are small and would not make visible contributions to $|\Delta_x|$ or $|\Delta_y|$ on the plots.

\begin{figure} 
\includegraphics[width=\hsize]{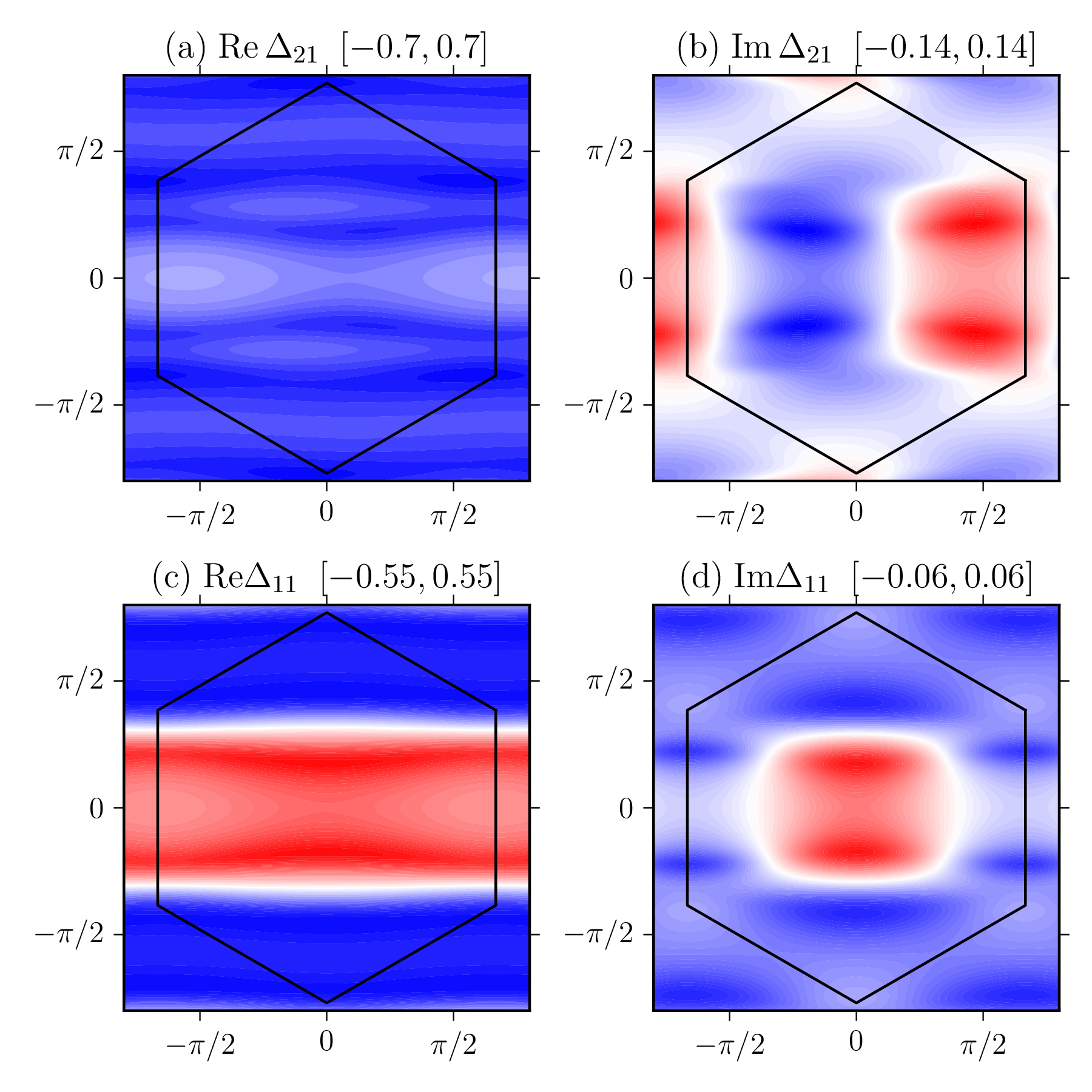}
\caption{(Color online) Superconducting order parameter $\Delta_{ab}(\kv)$ computed from the VCA solution, as a function of wave-vector. $U=10$ and doping is 10\%. This is to be compared with the BCS order parameter of Fig.~\ref{fig:gorkov_bcs}.}\label{fig:gorkov_C8}
\end{figure}

The order parameters shown on Fig.~\ref{fig:U10_C8-C12} are special convolutions of the general momentum-dependent order parameter
$\Delta_{ab}(\kv)$ with particular form factors associated with nearest-neighbor pairing. They have the advantage of simplicity, but are somewhat arbitrary. Unfortunately, the full order parameter $\Delta_{ab}(\kv)$ cannot be plotted simply as a function of doping. However, Figure~\ref{fig:gorkov_C8} shows $\Delta_{ab}(\kv)$ for the VCA solution at 10\% doping.
This is to be compared with Fig.~\ref{fig:gorkov_bcs}, which shows the corresponding BCS order parameter, obtained by setting the BCS fields to values that reproduce the same values of the link order parameters $\Delta_{x,y,\bot}$.
We notice that the features of Fig.~\ref{fig:gorkov_C8} are qualitatively the same as those of Fig.~\ref{fig:gorkov_bcs}, although less sharp, because of strong correlation effects. The sharp lines of Fig.~\ref{fig:gorkov_bcs} have become broad maxima and minima, but the asymmetry of $\Im\Delta_{21}$ stands out. Note that the scales (color range) differ from those of Fig.~\ref{fig:gorkov_bcs} by factors of two to three.

\begin{figure} 
\includegraphics[scale=0.85]{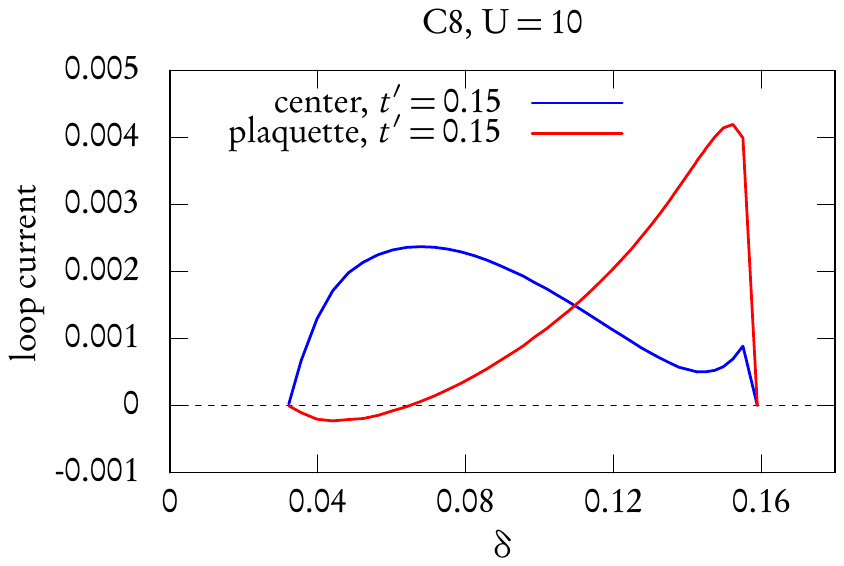}
\caption{(Color online) Loop supercurrents $I_p$ and $I_c$ circulating respectively around a plaquette and around the center parallelogram of the C8 cluster, as a function of doping, for $U=10$. The loops are illustrated on Fig.~\ref{fig:cluster}}      
\label{fig:C8_J}
\end{figure}

The $T$-breaking nature of the solutions found can also be assessed by computing chiral supercurrents.
Fig.~\ref{fig:C8_J} shows the supercurrents $I_p$ and $I_c$ circulating along the loops indicated on Fig.~\ref{fig:cluster}.
These are defined as the expectation values of 
\begin{equation}
\hat I = \frac1i \sum_{i\in\mathrm{loop},\sigma} \left(c_{i,\sigma}^\dagger c_{i+1,\sigma} - \Hc \right)
\end{equation}
where the sum is taken around the loop.
The expectation values $\langle\hat I_{p,c}\rangle_\mathrm{clus}$ in the cluster ground state vanishes if $\Delta_\bot$ is real, but is nonzero as soon as $\Delta_\bot$ develops an imaginary part.
This demonstrates that, in the latter case, superconductivity is chiral: if $\Im\Delta_\bot$ changes sign, the value of the Potthoff functional does not change -- hence we again have a VCA solution -- but the sign of the current changes.
Note that these supercurrents are measured on the cluster itself, as ground state expectation values, without using the Green function, because the latter provides expectation values on the whole lattice and these current loops cancel each other when the loops are stacked on the lattice. They are computed only to underline the chiral character of the complex superconducting solutions.
Note that the supercurrent loop located between the ladders dominates at small doping, whereas the contrary is true of the plaquette supercurrent.
This leads us to believe that the Cooper pairs tend to locate between the ladders at small doping and move towards the plaquettes at larger doping.

\begin{figure} 
\includegraphics[scale=0.85]{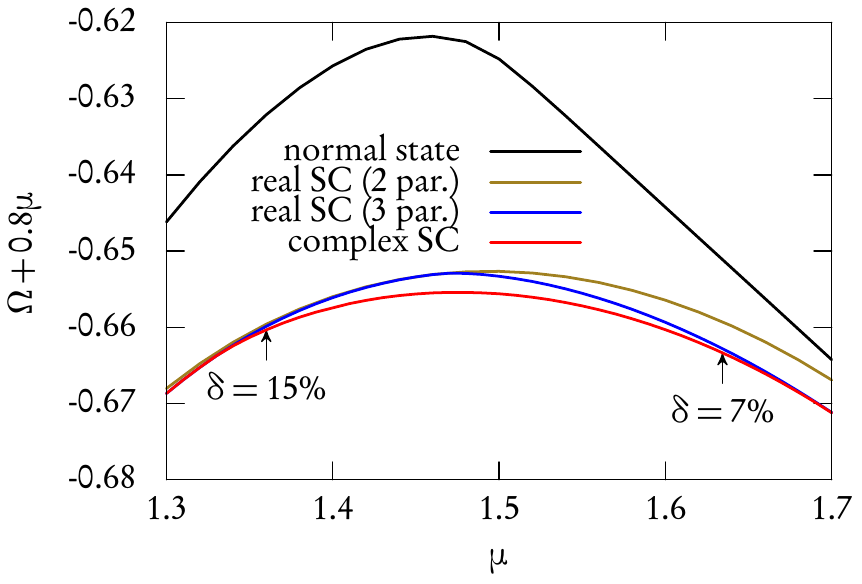}
\caption{(Color online) Value of the Potthoff functional at the solution as a function of chemical potential $\mu$ for $U=10$ and different sets of variational parameters: the normal solution (no variational parameters), two $T$-preserving solutions, with $(D_x,D_y)$ and $(D_x,D_y,\Re D_\bot)$ used as variational parameters, and the $T$-violating solution with variational set $(D_x,R_y,\Re D_\bot, \Im D_\bot)$. The latter has the lowest value of $\Omega$. Two values of doping are indicated for that solution.
A multiple of $\mu$ is added in order to better separate the different curves.
}      
\label{fig:cond}
\end{figure}

The $T$-breaking solution has the lowest energy in a sequence of solutions that can be obtained in VCA by increasing the number of variational parameters, as illustrated on Fig.~\ref{fig:cond}.
We plot the value of the Potthoff functional at the solution, which is an approximation to the grand potential $\Omega$, as a function of chemical potential $\mu$, since $\Omega$ is by construction as a function of $\mu$, not density.
In the top curve, no variational parameters were used, and this represents the normal solution. 
The second curve from the top is obtained by using the plaquette anomalous Weiss fields $(D_x,D_y)$ as variational parameters.
The third curve is obtained by adding the real part of the inter-ladder pairing $D_\bot$ to the set.
Finally, the lowest curve is obtained by adding both the real and imaginary parts of $D_\bot$ to the set, and the corresponding solutions break time-reversal invariance. This illustrates the process by which the quality of VCA solutions is improved by adding variational parameters. Another solution, obtained by allowing $D_y$ to take complex values, is not shown, as it is hardly distinguishable from the last one. We see that the energy advantage of the $T$-breaking solution is at best $0.0025t$, or roughly   $\frac1{15}$ of the condensation energy (the difference between the normal state and superconducting state energies), and this only at the most favorable doping ($\delta\sim 12\%$). Thus, even though the VCA simulation predicts $T$-breaking superconductivity in this system in a range of doping, it must be kept in mind that this solution is very close in energy to other approximate solutions that do not break time-reversal symmetry.

Notice that the difference between the second and third curves on Fig.~\ref{fig:cond} increases with $\mu$, i.e., towards smaller doping. This means that the importance of varying $D_\bot$ is greater on the underdoped side of the dome, which confirms our interpretation that the Cooper pairs tend to locate between the ladders at small doping.

\section{The Cellular Dynamical Mean Field Theory}\label{sec:cdmft}
We also used Cellular Dynamical Mean Field Theory (CDMFT) to confirm the appearance of a superconducting dome by independent means.

\subsection{Description of the method}

CDMFT like VCA, proceeds by tiling the lattice with clusters and by computing an optimized self-energy for each cluster. Unlike VCA, the space of self-energies is not explored by adding Weiss fields on the cluster, but rather by coupling each cluster to a bath of uncorrelated, auxiliary orbitals that represent the effect of the cluster's environment~\cite{Lichtenstein:2000vn,Kotliar:2001,Liebsch:2008mk,Senechal:2012kx}. The cluster Hamiltonian is supplemented by bath-cluster hybridization and bath energy terms: 
\begin{equation}\label{eq:Hbath}
 H_{\rm bath} = \sum_{\mu,\alpha} \theta_{\alpha\mu} a^\dagger_\mu c_\alpha  + 
 \sum_{\mu,\nu} \epsilon_{\mu\nu} a^\dagger_\mu a_\nu + \Hc
\end{equation}
where $a_\mu$ denotes the annihilation operator for the bath orbital labeled $\mu$.

This, together with the restriction of the Hubbard Hamiltonian \eqref{eq:H} to the cluster, defines an Anderson impurity model. The cluster Green function, when traced over the bath orbitals, takes the following form as a function of complex frequency $\om$:
\begin{equation}\label{eq:sigma}
\Gv'^{-1}(\om) = \om - \tv - \Gammav(\om) - \Sigmav(\om)
\end{equation}
where the hybridization matrix $\Gammav(\om)$ is
\begin{equation}\label{eq:hybridization}
\Gammav(\om) = \thetav(\om - \epsilonv)^{-1}\thetav^\dagger.
\end{equation}
in terms of the matrices $\theta_{\alpha\mu}$ and $\epsilon_{\mu\nu}$.
The Green function $\Gv(\kvt,\om)$ for the lattice model is then computed from the cluster's self-energy as
\begin{equation}
\Gv^{-1}(\kvt,\om) = \Gv_0^{-1}(\kvt,\om) - \Sigmav(\om)
\end{equation}
Here $\kvt$ denotes a reduced wave-vector, belonging to the Brillouin zone associated with the superlattice of clusters that defines the tiling. All Green function-related quantities are $2N_c\times 2N_c$ matrices, $N_c$ being the number of sites in the unit cell of the superlattice, which is made of one or more distinct clusters (the factor of 2 is there because of spin). $\Gv_0$ is the non-interacting Green function.
In practice, the cluster Green function is computed from an exact diagonalization technique using variants of the Lanczos method (just like in VCA). Then the self-energy is extracted from Eq.~\eqref{eq:sigma}.

\begin{figure} 
\includegraphics[height=6cm]{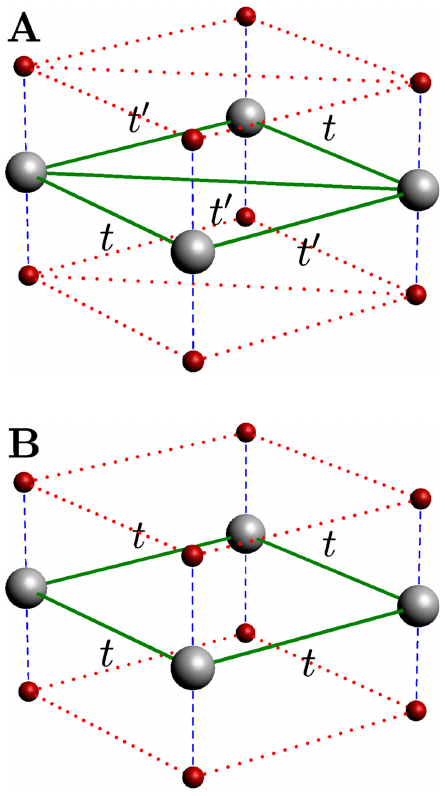}~\includegraphics[height=6cm]{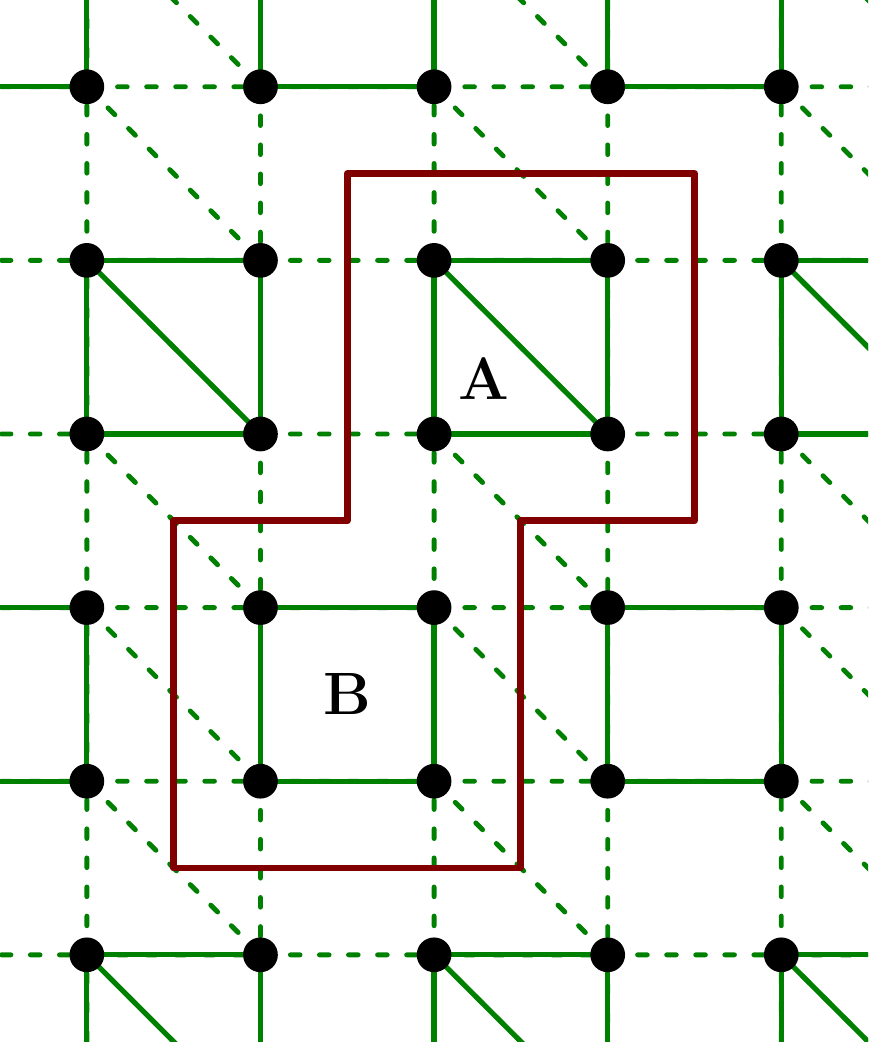}
\caption{(Color online) CDMFT cluster-bath system used in this work. 
The trellis lattice was deformed into a square lattice for simplicity.
Right panel: arrangement of the two clusters A and B needed to tile the lattice.  
Left panel: pictorial representation of the two clusters: lattice sites are gray spheres and bath orbitals are represented by smaller, red spheres. Green links represent hopping terms, dashed blue lines are bath-cluster hybridization terms and red dotted lines anomalous terms between bath sites, forming the anomalous part of the matrix $\epsilonv$.} 
\label{fig:2x2-8b}
\end{figure}

The bath and hybridization parameters $(\epsilon_{\mu\nu},\theta_{\alpha\mu})$ are determined by the self-consistency condition
\begin{equation}\label{eq:self-consistency}
\Gv'(\om) = \frac{N_c}N \sum_\kvt \Gv(\kvt,\om)
\end{equation} 
($N$ is the [quasi-infinite] number of sites in the whole system). In other words, the local Green function $\Gv'(\om)$ should coincide with the zero wave-vector Fourier transform of the full Green function. This condition should hold at all frequencies, which is impossible in a zero-temperature implementation of CDMFT because of the finite number of bath parameters at our disposal. Therefore, condition \eqref{eq:self-consistency} is only approximately satisfied, through the use of a merit function. Details can be found, for instance, in Ref.~\cite{Senechal:2012kx}.

\begin{figure} 
\includegraphics[width=\hsize]{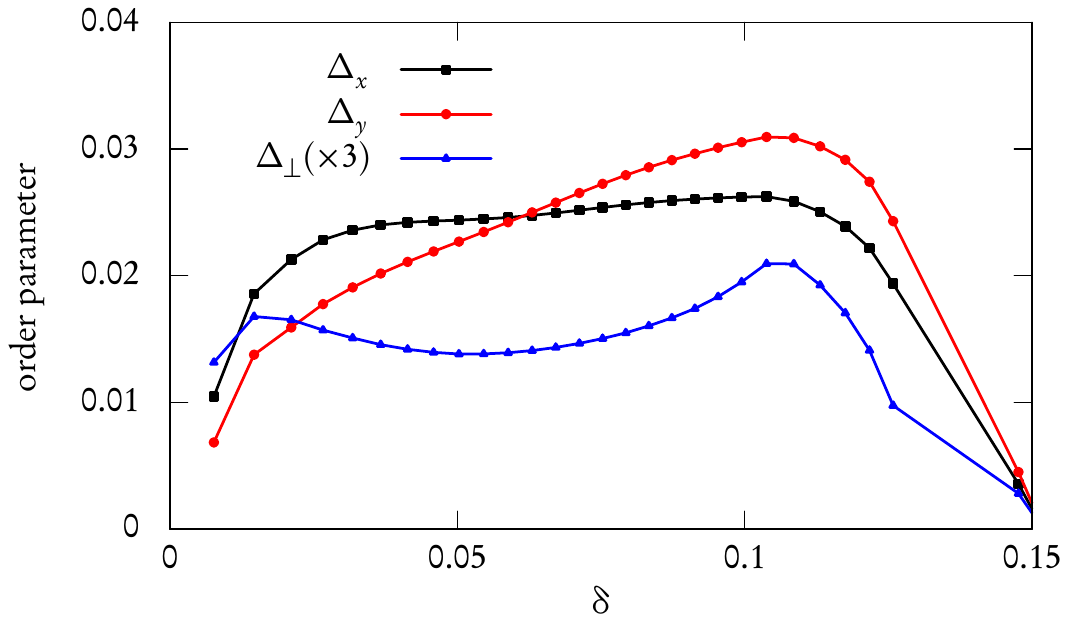}
\caption{(Color online) Order parameters computed from the CDMFT solutions found with the cluster-bath system illustrated in Fig.~\ref{fig:2x2-8b}, as a function of doping. The CDMFT solutions do not break time-reversal invariance.} 
\label{fig:cdmft}
\end{figure}

\subsection{System studied and superconductivity}

When modeling superconductivity in CDMFT, it is convenient to introduce anomalous terms between bath sites, thus treating bath sites as if they were forming a `phantom cluster'.
The cluster-bath system used in this work is illustrated on Fig.~\ref{fig:2x2-8b}.
Two unequivalent, four-site clusters form the repeated unit cell of the super-lattice.
The Nambu formalism is used to represent anomalous terms (see Ref.~\cite{Kancharla:2008vn} for explanations of its use in the context of CDMFT). The bath orbitals are grouped into two sets of four, and within each set anomalous terms are defined that mimic what could occur on the cluster itself (hence the expression `phantom cluster'). Each bath set has four or five links (dotted red lines on the figure) and a complex pairing operator is defined on each of these links, except on the rung link where it is assumed to be real, in order to set the global phase of the superconducting state. Taking symmetries into account, this makes for a total of 14 bath parameters for superconductivity, in addition to 8 bath orbital energies and as many hybridization parameters, for a total of 30 variational parameters.

Figure~\ref{fig:cdmft} shows the order parameters $\Delta_x$, $\Delta_y$ and $\Delta_\bot$ for the CDMFT solutions obtained at $U=10$.
These solutions do not break time-reversal in any significant way ($\Im\Delta_\bot < 10^{-5}$).
But the superconducting dome seen in VCA is still there, although somewhat narrower (nothing beyond $\delta=15\%$).
As doping is increased, the inter-ladder pairing operator $\Delta_\bot$ has a first maximum around 1.5\%, then decreases before increasing again, carried by the other components. This is another evidence that the Cooper pairs tend to gather between the ladders at small doping.

\begin{figure} 
\includegraphics[width=0.95\hsize]{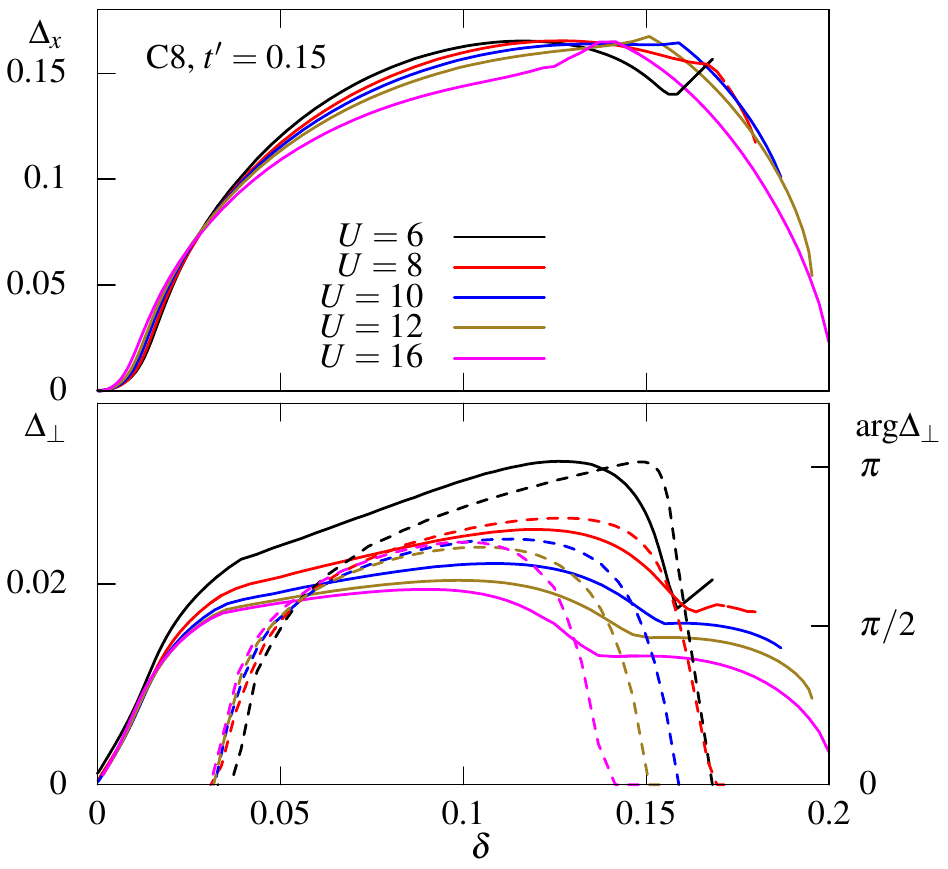}
\caption{(Color online) Order parameters $\Delta^x$ (top) and $|\Delta_\bot|$ (bottom) as a function of doping $\delta$ for cluster C8 and several values of on-site repulsion $U$.}      
\label{fig:C8_U}
\end{figure}

\section{Discussion}\label{sec:discussion}
Let us first point out an important difference between the present results, obtained for weakly coupled ladders, and superconductivity in the hole-doped, square lattice Hubbard model. In the latter~\cite{Senechal:2005, Kancharla:2008vn} the order parameter scales like $J\sim 4t^2/U$ at large $U$. Here, $\Delta_x$ is nearly $U$-independent in the range studied, as shown on Fig.~\ref{fig:C8_U}.
Changing the ratio $U/t$ is typically accomplished by applying pressure on the sample. However, in the case of
Sr$_{14-x}$Ca$_x$Cu$_{24}$O$_{41}$, changing the pressure would not only affect the value of $t$, but also of $t'$ and, more importantly, doping, as carriers migrate between the chains and the ladders. Thus mapping a change in $U$ to an experimentally accessible control parameter is very difficult.

Let us now discuss the origin of the chiral superconductivity that we have obtained in VCA.
It is known that in the repulsive, large-$U$ Hubbard model, the lattice symmetry and connectivity play an essential role in determining the symmetry of the order parameter.
On the square lattice, real $d$-wave $(d_{x^2-y^2})$ symmetry fits well with the four-fold coordination.
On the other hand, for triangular and honeycomb lattices, $d + id$ or chiral superconductivity fits well with the three- and six-fold coordination:
Chiral states carrying a $l_z=2$ angular momentum avoid nodes in the order parameter bond values in real space, thus gaining  condensation energy.
Likewise, in an isolated and isotropic ladder, $d$-wave symmetry fits well because of the plaquettes.
On the trellis lattice, we have elementary triangles, squares, and five-fold coordination.
A five-fold, odd number coordination in general accommodates a complex combination of $d$ and $s$ components.
Clearly the amplitude of the chiral component of superconductivity should increase with frustration, i.e., with $t'$.

How to explain, then, that VCA and CDMFT disagree on the chiral nature of superconductivity?
It may be that the small energy difference between the complex and real solutions shown on Fig.~\ref{fig:cond} cannot be resolved efficiently by CDMFT, but would be resolved if the same cluster-bath systems used in CDMFT were treated by Potthoff's self-energy functional approach (a method called CDIA); in practice, this is impossible to do because of the large number of variational parameters involved.

Despite this difference, the two approaches agree on important features: (1) The absence of superconductivity at half-filling: the system is then a Mott insulator; (2) the existence of a `dome' of $d$-wave superconductivity up to 15\% to 20\% doping; (3) the tendency of Cooper pairs to migrate from the inter-ladder regions to the plaquettes as doping is increased.
A careful study of the order parameter symmetry of the superconducting cuprate spin ladder compounds becomes important in the context of the possibility of chiral superconductivity found in this work.

\begin{acknowledgments}
Discussions with A.-M.S. Tremblay are gratefully acknowledged. Computing resources were provided by Compute Canada and Calcul Qu\'ebec.
\end{acknowledgments}


\begin{thebibliography}{24}%
\makeatletter
\providecommand \@ifxundefined [1]{%
 \@ifx{#1\undefined}
}%
\providecommand \@ifnum [1]{%
 \ifnum #1\expandafter \@firstoftwo
 \else \expandafter \@secondoftwo
 \fi
}%
\providecommand \@ifx [1]{%
 \ifx #1\expandafter \@firstoftwo
 \else \expandafter \@secondoftwo
 \fi
}%
\providecommand \natexlab [1]{#1}%
\providecommand \enquote  [1]{``#1''}%
\providecommand \bibnamefont  [1]{#1}%
\providecommand \bibfnamefont [1]{#1}%
\providecommand \citenamefont [1]{#1}%
\providecommand \href@noop [0]{\@secondoftwo}%
\providecommand \href [0]{\begingroup \@sanitize@url \@href}%
\providecommand \@href[1]{\@@startlink{#1}\@@href}%
\providecommand \@@href[1]{\endgroup#1\@@endlink}%
\providecommand \@sanitize@url [0]{\catcode `\\12\catcode `\$12\catcode
  `\&12\catcode `\#12\catcode `\^12\catcode `\_12\catcode `\%12\relax}%
\providecommand \@@startlink[1]{}%
\providecommand \@@endlink[0]{}%
\providecommand \url  [0]{\begingroup\@sanitize@url \@url }%
\providecommand \@url [1]{\endgroup\@href {#1}{\urlprefix }}%
\providecommand \urlprefix  [0]{URL }%
\providecommand \Eprint [0]{\href }%
\@ifxundefined \urlstyle {%
  \providecommand \doi  [0]{\begingroup \@sanitize@url \@doi}%
  \providecommand \@doi [1]{\endgroup \@@startlink {\doibase
  #1}doi:\discretionary {}{}{}#1\@@endlink }%
}{%
  \providecommand \doi  [0]{doi:\discretionary{}{}{}\begingroup
  \urlstyle{rm}\Url }%
}%
\providecommand \doibase [0]{http://dx.doi.org/}%
\providecommand \Doi [0]{\begingroup \@sanitize@url \@Doi }%
\providecommand \@Doi  [1]{\endgroup\@@startlink{\doibase#1}\@@Doi}%
\providecommand \@@Doi [1]{#1\@@endlink}%
\providecommand \selectlanguage [0]{\@gobble}%
\providecommand \bibinfo  [0]{\@secondoftwo}%
\providecommand \bibfield  [0]{\@secondoftwo}%
\providecommand \translation [1]{[#1]}%
\providecommand \BibitemOpen [0]{}%
\providecommand \bibitemStop [0]{}%
\providecommand \bibitemNoStop [0]{.\EOS\space}%
\providecommand \EOS [0]{\spacefactor3000\relax}%
\providecommand \BibitemShut  [1]{\csname bibitem#1\endcsname}%
\bibitem [{\citenamefont {Dagotto}\ and\ \citenamefont
  {Moreo}(1988)}]{Dagotto:1988}%
  \BibitemOpen
  \bibfield  {author} {\bibinfo {author} {\bibfnamefont {E.}~\bibnamefont
  {Dagotto}}\ and\ \bibinfo {author} {\bibfnamefont {A.}~\bibnamefont
  {Moreo}},\ }\href {http://link.aps.org/doi/10.1103/PhysRevB.38.5087}
  {\bibfield  {journal} {\bibinfo  {journal} {Phys. Rev. B},\ }\textbf
  {\bibinfo {volume} {38}},\ \bibinfo {pages} {5087} (\bibinfo {year}
  {1988})}\BibitemShut {NoStop}%
\bibitem [{\citenamefont {Dagotto}\ \emph {et~al.}(1992)\citenamefont
  {Dagotto}, \citenamefont {Riera},\ and\ \citenamefont
  {Scalapino}}]{Dagotto:1992}%
  \BibitemOpen
  \bibfield  {author} {\bibinfo {author} {\bibfnamefont {E.}~\bibnamefont
  {Dagotto}}, \bibinfo {author} {\bibfnamefont {J.}~\bibnamefont {Riera}}, \
  and\ \bibinfo {author} {\bibfnamefont {D.}~\bibnamefont {Scalapino}},\ }\Doi
  {10.1103/PhysRevB.45.5744} {\bibfield  {journal} {\bibinfo  {journal} {Phys.
  Rev. B},\ }\textbf {\bibinfo {volume} {45}},\ \bibinfo {pages} {5744}
  (\bibinfo {year} {1992})}\BibitemShut {NoStop}%
\bibitem [{\citenamefont {Barnes}\ \emph {et~al.}(1993)\citenamefont {Barnes},
  \citenamefont {Dagotto}, \citenamefont {Riera},\ and\ \citenamefont
  {Swanson}}]{Barnes:1993}%
  \BibitemOpen
  \bibfield  {author} {\bibinfo {author} {\bibfnamefont {T.}~\bibnamefont
  {Barnes}}, \bibinfo {author} {\bibfnamefont {E.}~\bibnamefont {Dagotto}},
  \bibinfo {author} {\bibfnamefont {J.}~\bibnamefont {Riera}}, \ and\ \bibinfo
  {author} {\bibfnamefont {E.~S.}\ \bibnamefont {Swanson}},\ }\Doi
  {10.1103/PhysRevB.47.3196} {\bibfield  {journal} {\bibinfo  {journal} {Phys.
  Rev. B},\ }\textbf {\bibinfo {volume} {47}},\ \bibinfo {pages} {3196}
  (\bibinfo {year} {1993})}\BibitemShut {NoStop}%
\bibitem [{\citenamefont {Uehara}\ \emph {et~al.}(1996)\citenamefont {Uehara},
  \citenamefont {Nagata}, \citenamefont {Akimitsu}, \citenamefont {Takahashi},
  \citenamefont {M{\^o}ri},\ and\ \citenamefont {Kinoshita}}]{Uehara:1996}%
  \BibitemOpen
  \bibfield  {author} {\bibinfo {author} {\bibfnamefont {M.}~\bibnamefont
  {Uehara}}, \bibinfo {author} {\bibfnamefont {T.}~\bibnamefont {Nagata}},
  \bibinfo {author} {\bibfnamefont {J.}~\bibnamefont {Akimitsu}}, \bibinfo
  {author} {\bibfnamefont {H.}~\bibnamefont {Takahashi}}, \bibinfo {author}
  {\bibfnamefont {N.}~\bibnamefont {M{\^o}ri}}, \ and\ \bibinfo {author}
  {\bibfnamefont {K.}~\bibnamefont {Kinoshita}},\ }\href@noop {} {\bibfield
  {journal} {\bibinfo  {journal} {Journal of the Physical Society of Japan},\
  }\textbf {\bibinfo {volume} {65}},\ \bibinfo {pages} {2764} (\bibinfo {year}
  {1996})}\BibitemShut {NoStop}%
\bibitem [{\citenamefont {Nagata}\ \emph {et~al.}(1997)\citenamefont {Nagata},
  \citenamefont {Uehara}, \citenamefont {Goto}, \citenamefont {Komiya},
  \citenamefont {Akimitsu}, \citenamefont {Motoyama}, \citenamefont {Eisaki},
  \citenamefont {Uchida}, \citenamefont {Takahashi}, \citenamefont {Nakanishi}
  \emph {et~al.}}]{Nagata:1997}%
  \BibitemOpen
  \bibfield  {author} {\bibinfo {author} {\bibfnamefont {T.}~\bibnamefont
  {Nagata}}, \bibinfo {author} {\bibfnamefont {M.}~\bibnamefont {Uehara}},
  \bibinfo {author} {\bibfnamefont {J.}~\bibnamefont {Goto}}, \bibinfo {author}
  {\bibfnamefont {N.}~\bibnamefont {Komiya}}, \bibinfo {author} {\bibfnamefont
  {J.}~\bibnamefont {Akimitsu}}, \bibinfo {author} {\bibfnamefont
  {N.}~\bibnamefont {Motoyama}}, \bibinfo {author} {\bibfnamefont
  {H.}~\bibnamefont {Eisaki}}, \bibinfo {author} {\bibfnamefont
  {S.}~\bibnamefont {Uchida}}, \bibinfo {author} {\bibfnamefont
  {H.}~\bibnamefont {Takahashi}}, \bibinfo {author} {\bibfnamefont
  {T.}~\bibnamefont {Nakanishi}},  \emph {et~al.},\ }\href@noop {} {\bibfield
  {journal} {\bibinfo  {journal} {Physica C: Superconductivity},\ }\textbf
  {\bibinfo {volume} {282}},\ \bibinfo {pages} {153} (\bibinfo {year}
  {1997})}\BibitemShut {NoStop}%
\bibitem [{\citenamefont {{Mohan Radheep}}\ \emph {et~al.}(2013)\citenamefont
  {{Mohan Radheep}}, \citenamefont {{Thiyagarjan}}, \citenamefont
  {{Esakkimuthu}}, \citenamefont {{Deng}}, \citenamefont {{Pomjakushina}},
  \citenamefont {{Prajapat}}, \citenamefont {{Ravikumar}}, \citenamefont
  {{Conder}}, \citenamefont {{Baskaran}},\ and\ \citenamefont
  {{Arumugam}}}]{Mohan-Radheep:2013kq}%
  \BibitemOpen
  \bibfield  {author} {\bibinfo {author} {\bibfnamefont {D.}~\bibnamefont
  {{Mohan Radheep}}}, \bibinfo {author} {\bibfnamefont {R.}~\bibnamefont
  {{Thiyagarjan}}}, \bibinfo {author} {\bibfnamefont {S.}~\bibnamefont
  {{Esakkimuthu}}}, \bibinfo {author} {\bibfnamefont {G.}~\bibnamefont
  {{Deng}}}, \bibinfo {author} {\bibfnamefont {E.}~\bibnamefont
  {{Pomjakushina}}}, \bibinfo {author} {\bibfnamefont {C.~L.}\ \bibnamefont
  {{Prajapat}}}, \bibinfo {author} {\bibfnamefont {G.}~\bibnamefont
  {{Ravikumar}}}, \bibinfo {author} {\bibfnamefont {K.}~\bibnamefont
  {{Conder}}}, \bibinfo {author} {\bibfnamefont {G.}~\bibnamefont
  {{Baskaran}}}, \ and\ \bibinfo {author} {\bibfnamefont {S.}~\bibnamefont
  {{Arumugam}}},\ }\href@noop {} {\bibfield  {journal} {\bibinfo  {journal}
  {ArXiv e-prints}} (\bibinfo {year} {2013})},\ \Eprint
  {http://arxiv.org/abs/1303.0921} {arXiv:1303.0921 [cond-mat.supr-con]}
  \BibitemShut {NoStop}%
\bibitem [{\citenamefont {Troyer}\ \emph {et~al.}(1996)\citenamefont {Troyer},
  \citenamefont {Tsunetsugu},\ and\ \citenamefont {Rice}}]{Troyer:1996fk}%
  \BibitemOpen
  \bibfield  {author} {\bibinfo {author} {\bibfnamefont {M.}~\bibnamefont
  {Troyer}}, \bibinfo {author} {\bibfnamefont {H.}~\bibnamefont {Tsunetsugu}},
  \ and\ \bibinfo {author} {\bibfnamefont {T.~M.}\ \bibnamefont {Rice}},\
  }\href {http://link.aps.org/doi/10.1103/PhysRevB.53.251} {\bibfield
  {journal} {\bibinfo  {journal} {Phys. Rev. B},\ }\textbf {\bibinfo {volume}
  {53}},\ \bibinfo {pages} {251} (\bibinfo {year} {1996})}\BibitemShut
  {NoStop}%
\bibitem [{\citenamefont {Noack}\ \emph {et~al.}(1994)\citenamefont {Noack},
  \citenamefont {White},\ and\ \citenamefont {Scalapino}}]{Noack:1994}%
  \BibitemOpen
  \bibfield  {author} {\bibinfo {author} {\bibfnamefont {R.~M.}\ \bibnamefont
  {Noack}}, \bibinfo {author} {\bibfnamefont {S.~R.}\ \bibnamefont {White}}, \
  and\ \bibinfo {author} {\bibfnamefont {D.~J.}\ \bibnamefont {Scalapino}},\
  }\href {http://link.aps.org/doi/10.1103/PhysRevLett.73.882} {\bibfield
  {journal} {\bibinfo  {journal} {Phys. Rev. Lett.},\ }\textbf {\bibinfo
  {volume} {73}},\ \bibinfo {pages} {882} (\bibinfo {year} {1994})}\BibitemShut
  {NoStop}%
\bibitem [{\citenamefont {Gopalan}\ \emph {et~al.}(1994)\citenamefont
  {Gopalan}, \citenamefont {Rice},\ and\ \citenamefont
  {Sigrist}}]{Gopalan:1994}%
  \BibitemOpen
  \bibfield  {author} {\bibinfo {author} {\bibfnamefont {S.}~\bibnamefont
  {Gopalan}}, \bibinfo {author} {\bibfnamefont {T.~M.}\ \bibnamefont {Rice}}, \
  and\ \bibinfo {author} {\bibfnamefont {M.}~\bibnamefont {Sigrist}},\ }\href
  {http://link.aps.org/doi/10.1103/PhysRevB.49.8901} {\bibfield  {journal}
  {\bibinfo  {journal} {Phys. Rev. B},\ }\textbf {\bibinfo {volume} {49}},\
  \bibinfo {pages} {8901} (\bibinfo {year} {1994})}\BibitemShut {NoStop}%
\bibitem [{\citenamefont {Balents}\ and\ \citenamefont
  {Fisher}(1996)}]{Balents:1996}%
  \BibitemOpen
  \bibfield  {author} {\bibinfo {author} {\bibfnamefont {L.}~\bibnamefont
  {Balents}}\ and\ \bibinfo {author} {\bibfnamefont {M.~P.~A.}\ \bibnamefont
  {Fisher}},\ }\href {http://link.aps.org/doi/10.1103/PhysRevB.53.12133}
  {\bibfield  {journal} {\bibinfo  {journal} {Phys. Rev. B},\ }\textbf
  {\bibinfo {volume} {53}},\ \bibinfo {pages} {12133} (\bibinfo {year}
  {1996})}\BibitemShut {NoStop}%
\bibitem [{\citenamefont {Kuroki}\ \emph {et~al.}(1996)\citenamefont {Kuroki},
  \citenamefont {Kimura},\ and\ \citenamefont {Aoki}}]{Kuroki:1996}%
  \BibitemOpen
  \bibfield  {author} {\bibinfo {author} {\bibfnamefont {K.}~\bibnamefont
  {Kuroki}}, \bibinfo {author} {\bibfnamefont {T.}~\bibnamefont {Kimura}}, \
  and\ \bibinfo {author} {\bibfnamefont {H.}~\bibnamefont {Aoki}},\ }\href
  {http://link.aps.org/doi/10.1103/PhysRevB.54.R15641} {\bibfield  {journal}
  {\bibinfo  {journal} {Phys. Rev. B},\ }\textbf {\bibinfo {volume} {54}},\
  \bibinfo {pages} {R15641} (\bibinfo {year} {1996})}\BibitemShut {NoStop}%
\bibitem [{\citenamefont {Dahm}\ and\ \citenamefont
  {Scalapino}(1997)}]{Dahm:1997}%
  \BibitemOpen
  \bibfield  {author} {\bibinfo {author} {\bibfnamefont {T.}~\bibnamefont
  {Dahm}}\ and\ \bibinfo {author} {\bibfnamefont {D.}~\bibnamefont
  {Scalapino}},\ }\href@noop {} {\bibfield  {journal} {\bibinfo  {journal}
  {Physica C: Superconductivity},\ }\textbf {\bibinfo {volume} {288}},\
  \bibinfo {pages} {33} (\bibinfo {year} {1997})}\BibitemShut {NoStop}%
\bibitem [{\citenamefont {Kontani}\ and\ \citenamefont
  {Ueda}(1998)}]{Kontani:1998}%
  \BibitemOpen
  \bibfield  {author} {\bibinfo {author} {\bibfnamefont {H.}~\bibnamefont
  {Kontani}}\ and\ \bibinfo {author} {\bibfnamefont {K.}~\bibnamefont {Ueda}},\
  }\href@noop {} {\bibfield  {journal} {\bibinfo  {journal} {Physical review
  letters},\ }\textbf {\bibinfo {volume} {80}},\ \bibinfo {pages} {5619}
  (\bibinfo {year} {1998})}\BibitemShut {NoStop}%
\bibitem [{\citenamefont {Dahnken}\ \emph {et~al.}(2004)\citenamefont
  {Dahnken}, \citenamefont {Aichhorn}, \citenamefont {Hanke}, \citenamefont
  {Arrigoni},\ and\ \citenamefont {Potthoff}}]{Dahnken:2004}%
  \BibitemOpen
  \bibfield  {author} {\bibinfo {author} {\bibfnamefont {C.}~\bibnamefont
  {Dahnken}}, \bibinfo {author} {\bibfnamefont {M.}~\bibnamefont {Aichhorn}},
  \bibinfo {author} {\bibfnamefont {W.}~\bibnamefont {Hanke}}, \bibinfo
  {author} {\bibfnamefont {E.}~\bibnamefont {Arrigoni}}, \ and\ \bibinfo
  {author} {\bibfnamefont {M.}~\bibnamefont {Potthoff}},\ }\href
  {http://link.aps.org/doi/10.1103/PhysRevB.70.245110} {\bibfield  {journal}
  {\bibinfo  {journal} {Phys. Rev. B},\ }\textbf {\bibinfo {volume} {70}},\
  \bibinfo {pages} {245110} (\bibinfo {year} {2004})}\BibitemShut {NoStop}%
\bibitem [{\citenamefont {Lichtenstein}\ and\ \citenamefont
  {Katsnelson}(2000)}]{Lichtenstein:2000vn}%
  \BibitemOpen
  \bibfield  {author} {\bibinfo {author} {\bibfnamefont {A.~I.}\ \bibnamefont
  {Lichtenstein}}\ and\ \bibinfo {author} {\bibfnamefont {M.~I.}\ \bibnamefont
  {Katsnelson}},\ }\Doi {10.1103/PhysRevB.62.R9283} {\bibfield  {journal}
  {\bibinfo  {journal} {Phys. Rev. B},\ }\textbf {\bibinfo {volume} {62}},\
  \bibinfo {pages} {R9283} (\bibinfo {year} {2000})}\BibitemShut {NoStop}%
\bibitem [{\citenamefont {Kotliar}\ \emph {et~al.}(2001)\citenamefont
  {Kotliar}, \citenamefont {Savrasov}, \citenamefont {P{\'a}lsson},\ and\
  \citenamefont {Biroli}}]{Kotliar:2001}%
  \BibitemOpen
  \bibfield  {author} {\bibinfo {author} {\bibfnamefont {G.}~\bibnamefont
  {Kotliar}}, \bibinfo {author} {\bibfnamefont {S.}~\bibnamefont {Savrasov}},
  \bibinfo {author} {\bibfnamefont {G.}~\bibnamefont {P{\'a}lsson}}, \ and\
  \bibinfo {author} {\bibfnamefont {G.}~\bibnamefont {Biroli}},\ }\href@noop {}
  {\bibfield  {journal} {\bibinfo  {journal} {Phys. Rev. Lett.},\ }\textbf
  {\bibinfo {volume} {87}},\ \bibinfo {pages} {186401} (\bibinfo {year}
  {2001})}\BibitemShut {NoStop}%
\bibitem [{\citenamefont {Arai}\ and\ \citenamefont
  {Tsunetsugu}(1997)}]{Arai:1997}%
  \BibitemOpen
  \bibfield  {author} {\bibinfo {author} {\bibfnamefont {M.}~\bibnamefont
  {Arai}}\ and\ \bibinfo {author} {\bibfnamefont {H.}~\bibnamefont
  {Tsunetsugu}},\ }\href {http://link.aps.org/doi/10.1103/PhysRevB.56.R4305}
  {\bibfield  {journal} {\bibinfo  {journal} {Phys. Rev. B},\ }\textbf
  {\bibinfo {volume} {56}},\ \bibinfo {pages} {R4305} (\bibinfo {year}
  {1997})}\BibitemShut {NoStop}%
\bibitem [{\citenamefont {S{\'e}n{\'e}chal}\ \emph {et~al.}(2005)\citenamefont
  {S{\'e}n{\'e}chal}, \citenamefont {Lavertu}, \citenamefont {Marois},\ and\
  \citenamefont {Tremblay}}]{Senechal:2005}%
  \BibitemOpen
  \bibfield  {author} {\bibinfo {author} {\bibfnamefont {D.}~\bibnamefont
  {S{\'e}n{\'e}chal}}, \bibinfo {author} {\bibfnamefont {P.-L.}\ \bibnamefont
  {Lavertu}}, \bibinfo {author} {\bibfnamefont {M.-A.}\ \bibnamefont {Marois}},
  \ and\ \bibinfo {author} {\bibfnamefont {A.-M.~S.}\ \bibnamefont
  {Tremblay}},\ }\href {http://link.aps.org/doi/10.1103/PhysRevLett.94.156404}
  {\bibfield  {journal} {\bibinfo  {journal} {Phys. Rev. Lett.},\ }\textbf
  {\bibinfo {volume} {94}},\ \bibinfo {pages} {156404} (\bibinfo {year}
  {2005})}\BibitemShut {NoStop}%
\bibitem [{\citenamefont {Aichhorn}\ \emph {et~al.}(2006)\citenamefont
  {Aichhorn}, \citenamefont {Arrigoni}, \citenamefont {Potthoff},\ and\
  \citenamefont {Hanke}}]{Aichhorn:2006rt}%
  \BibitemOpen
  \bibfield  {author} {\bibinfo {author} {\bibfnamefont {M.}~\bibnamefont
  {Aichhorn}}, \bibinfo {author} {\bibfnamefont {E.}~\bibnamefont {Arrigoni}},
  \bibinfo {author} {\bibfnamefont {M.}~\bibnamefont {Potthoff}}, \ and\
  \bibinfo {author} {\bibfnamefont {W.}~\bibnamefont {Hanke}},\ }\Doi
  {10.1103/PhysRevB.74.235117} {\bibfield  {journal} {\bibinfo  {journal}
  {Phys. Rev. B},\ }\textbf {\bibinfo {volume} {74}},\ \bibinfo {eid} {235117}
  (\bibinfo {year} {2006})}\BibitemShut {NoStop}%
\bibitem [{\citenamefont {Potthoff}(2003)}]{Potthoff:2003b}%
  \BibitemOpen
  \bibfield  {author} {\bibinfo {author} {\bibfnamefont {M.}~\bibnamefont
  {Potthoff}},\ }\Doi {10.1140/epjb/e2003-00121-8} {\bibfield  {journal}
  {\bibinfo  {journal} {Eur. Phys. J. B},\ }\textbf {\bibinfo {volume} {32}},\
  \bibinfo {pages} {429 } (\bibinfo {year} {2003})}\BibitemShut {NoStop}%
\bibitem [{\citenamefont {Potthoff}(2012)}]{Potthoff:2012}%
  \BibitemOpen
  \bibfield  {author} {\bibinfo {author} {\bibfnamefont {M.}~\bibnamefont
  {Potthoff}},\ }in\ \href@noop {} {\emph {\bibinfo {booktitle} {Theoretical
  methods for Strongly Correlated Systems}}},\ \bibinfo {series} {Springer
  Series in Solid-State Sciences}, Vol.\ \bibinfo {volume} {171},\ \bibinfo
  {editor} {edited by\ \bibinfo {editor} {\bibfnamefont {A.}~\bibnamefont
  {Avella}}\ and\ \bibinfo {editor} {\bibfnamefont {F.}~\bibnamefont
  {Mancini}}}\ (\bibinfo  {publisher} {Springer},\ \bibinfo {year} {2012})\
  Chap.~\bibinfo {chapter} {9}\BibitemShut {NoStop}%
\bibitem [{\citenamefont {Liebsch}\ \emph {et~al.}(2008)\citenamefont
  {Liebsch}, \citenamefont {Ishida},\ and\ \citenamefont
  {Merino}}]{Liebsch:2008mk}%
  \BibitemOpen
  \bibfield  {author} {\bibinfo {author} {\bibfnamefont {A.}~\bibnamefont
  {Liebsch}}, \bibinfo {author} {\bibfnamefont {H.}~\bibnamefont {Ishida}}, \
  and\ \bibinfo {author} {\bibfnamefont {J.}~\bibnamefont {Merino}},\ }\Doi
  {10.1103/PhysRevB.78.165123} {\bibfield  {journal} {\bibinfo  {journal}
  {Phys. Rev. B},\ }\textbf {\bibinfo {volume} {78}},\ \bibinfo {pages}
  {165123} (\bibinfo {year} {2008})}\BibitemShut {NoStop}%
\bibitem [{\citenamefont {S{\'e}n{\'e}chal}(2012)}]{Senechal:2012kx}%
  \BibitemOpen
  \bibfield  {author} {\bibinfo {author} {\bibfnamefont {D.}~\bibnamefont
  {S{\'e}n{\'e}chal}},\ }in\ \href
  {http://dx.doi.org/10.1007/978-3-642-21831-6_11} {\emph {\bibinfo {booktitle}
  {Strongly Correlated Systems}}},\ \bibinfo {series} {Springer Series in
  Solid-State Sciences}, Vol.\ \bibinfo {volume} {171},\ \bibinfo {editor}
  {edited by\ \bibinfo {editor} {\bibfnamefont {A.}~\bibnamefont {Avella}}\
  and\ \bibinfo {editor} {\bibfnamefont {F.}~\bibnamefont {Mancini}}}\
  (\bibinfo  {publisher} {Springer Berlin Heidelberg},\ \bibinfo {year}
  {2012})\ pp.\ \bibinfo {pages} {341--371},\ ISBN \bibinfo {isbn}
  {978-3-642-21831-6}\BibitemShut {NoStop}%
\bibitem [{\citenamefont {Kancharla}\ \emph {et~al.}(2008)\citenamefont
  {Kancharla}, \citenamefont {Kyung}, \citenamefont {S{\'e}n{\'e}chal},
  \citenamefont {Civelli}, \citenamefont {Capone}, \citenamefont {Kotliar},\
  and\ \citenamefont {Tremblay}}]{Kancharla:2008vn}%
  \BibitemOpen
  \bibfield  {author} {\bibinfo {author} {\bibfnamefont {S.~S.}\ \bibnamefont
  {Kancharla}}, \bibinfo {author} {\bibfnamefont {B.}~\bibnamefont {Kyung}},
  \bibinfo {author} {\bibfnamefont {D.}~\bibnamefont {S{\'e}n{\'e}chal}},
  \bibinfo {author} {\bibfnamefont {M.}~\bibnamefont {Civelli}}, \bibinfo
  {author} {\bibfnamefont {M.}~\bibnamefont {Capone}}, \bibinfo {author}
  {\bibfnamefont {G.}~\bibnamefont {Kotliar}}, \ and\ \bibinfo {author}
  {\bibfnamefont {A.-M.~S.}\ \bibnamefont {Tremblay}},\ }\Doi
  {10.1103/PhysRevB.77.184516} {\bibfield  {journal} {\bibinfo  {journal}
  {Phys. Rev. B},\ }\textbf {\bibinfo {volume} {77}},\ \bibinfo {pages}
  {184516} (\bibinfo {year} {2008})}\BibitemShut {NoStop}%
\end{thebibliography}

%

\end{document}